\documentclass[prd,superscriptaddress,showpacs,showkeys,twocolumn]{revtex4}
\usepackage[T1]{fontenc}
\usepackage{amsmath}
\usepackage{amssymb}
\usepackage[colorinlistoftodos]{todonotes}
\usepackage{epsfig}
\usepackage{changes}
\usepackage{palatino}
\def\be{\begin{equation}}
\def\ee{\end{equation}}
\usepackage[colorlinks=true,
            linkcolor=blue,
           urlcolor=blue,
           citecolor=blue]{hyperref}
\usepackage{tabstackengine}
\usepackage{graphicx}
\usepackage[english]{babel}
\usepackage{amsfonts}
\usepackage{eurosym}
\usepackage{calrsfs}
\usepackage{color}
\usepackage{float}
\begin{document}

\title{Weak Gravitational lensing by phantom black holes and phantom wormholes using the Gauss-Bonnet theorem}

\author{Ali \"{O}vg\"{u}n}
\email{ali.ovgun@pucv.cl}

\affiliation{Instituto de F\'{\i}sica, Pontificia Universidad Cat\'olica de
Valpara\'{\i}so, Casilla 4950, Valpara\'{\i}so, Chile.}

\affiliation{Physics Department, Arts and Sciences Faculty, Eastern Mediterranean University, Famagusta, North Cyprus via Mersin 10, Turkey.}

\author{Galin Gyulchev}

\email{gyulchev@phys.uni-sofia.bg}

\affiliation{Faculty of Physics, St. Kliment Ohridski University of Sofia, 5 James Bourchier Boulevard, Sofia 1164, Bulgaria}

\affiliation{Department of Physics, Biophysics and Roentgenology, Faculty of Medicine, St. Kliment Ohridski University of Sofia, 1, Kozyak Str., 1407 Sofia, Bulgaria.}

\author{Kimet Jusufi}
\email{kimet.jusufi@unite.edu.mk}
\affiliation{Physics Department, State University of Tetovo, Ilinden Street nn, 1200,
Tetovo, North Macedonia.}
\affiliation{Institute of Physics, Faculty of Natural Sciences and Mathematics, Ss. Cyril and Methodius University, Arhimedova 3, 1000 Skopje, North Macedonia.}

\begin{abstract}
In this paper, we  study the deflection of light by a class of phantom black hole and wormhole solutions in the weak limit approximation. More specifically, in the first part of this work we study the deflection of light by Garfinkle-Horowitz-Str\"{o}minger black hole and Einstein-Maxwell anti-dilaton black hole using the optical geometry and the Gauss-Bonnet theorem. Our calculation show that gravitational lensing is affected by the phantom scalar field (phantom dilaton). In the second part of this work, we explore the deflection of light by a class of asymptotically flat phantom wormholes. In particular we have used three types of wormholes: wormhole with a bounded/unbounded mass function, and a wormhole with a vanishing redshift function. We show that the particular choice of the shape function and mass function plays a crucial role in the final expression for the deflection angle of light. In the third part of the paper we verify our findings with the help of standard geodesics equations. Finally, in the fourth part of this paper we consider the problem for the observational relevance of our results studying the creation of the weak field Einstein rings.
\end{abstract}

\keywords{Gravitational lensing; Phantom black holes; Einstein-Maxwell-dilaton theory; Wormholes}

\pacs{95.30.Sf, 04.20.Dw, 04.70.Bw, 98.62.Sb}
\date{\today}

\maketitle

\section{Introduction}

Presently, the general relativity is not the only viable theory of gravitation. One of the most promising avenue realizing the unified theories is the string theory, which in the low-energy limit reduces to the Einstein-Maxwell dilaton gravity. Moreover, the most important questions nowadays is devoted to the existence of the dark energy (DE). The natural questions arises whether local manifestations of DE at astrophysical scale can be observed, as considerable efforts are made to study the nature of DE. Different effective models of dark energy have been proposed in literature \cite{Miao,Tsujikawa}. In some of them the possibility of describing DE by phantom fields is considered.

Moreover, one of the modified gravity theory for explaining the dark energy is the Einstein--(anti--) Maxwell dilaton theory (EMDT), which has both Maxwell fields and a phantom dilatonic field with a "wrong" sign of kinetic energy. In the EMDT, various type of black holes were found and discussed in \cite{Bronnikov:1977is,Bolokhov:2012kn,Gibbons:1982ih,Gibbons:1987ps,Gurses:1995fw,Lazov:2019bni,Wang:2018civ,Rocha:2018lmv,Khalil:2018aaj,Azreg-Ainou:2018wjx,Pacilio:2018gom,Azreg-Ainou:2018tkl}. First, Gibbons and Rasheed obtained the phantom black hole solutions using the EMDT \cite{Gibbons:1996pd}. Afterwards, the solution of phantom black hole was generalized in higher dimensions by Clement et al. \cite{Clement:2009ai,AzregAinou:2011rj}. Next, the solution of the regular phantom black holes were obtained \cite{Bronnikov:2005gm,Rodrigues:2012tw}.

The gravitational lensing can be used for a confirmation of the generalized gravity theories. In most of the cases of interest connected with phenomena realizing on large scales, it is possible to assume that the gravitational field is weak, hence the angle of deflection of the light rays caused by a spherically symmetric body with mass $M$  can be approximated by:
$\hat{\alpha} \approx 4GM/bc^{2}$, where $b$ is the impact parameter. But the question of the influence of the phantom black hole charge on the light deflection angle in the weak field regime still remains unclear.

As we have already mentioned, the weak gravitational lensing could provide examination in the asymptotically flat space-time region of different kinds black holes. Therefore, following the method of Gibbons and Werner (GW) in the present work we wish to study gravitational lensing in the weak field limit due to a static, spherically symmetric, charged dilaton Garfinkle-Horowitz-Strominger (GHS) black hole \cite{Garfinkle:1990qj} in the heterotic string theory with the aim of investigating the influence of the dilaton field in the behaviour of the light bending angle in the weak deflection limit regime by employing the Gauss-Bonnet theorem (GBT) to the optical geometry. Nowadays, GBT has been applied in numerous studies
\cite{Ovgun:2018xys,asahi1,Ono:2017pie,  Jusufiwh,Jusufi:2017mav,kimet2,kimet22,wh10,kimet1,kaa,kimet3,aovgun,aovgun1,asada,Crisnejo:2018uyn,Jusufi:2018jof,Jusufi:2018kmk,Jusufi:2017drg,Jusufi:2017uhh,Jusufi:2017xnr,Jusufi:2017vew,Ishihara:2016vdc,Jusufi:2015laa,Gibbons:2015qja}.

Deflection of light normally depends on the mass of the enclosed region, however for the GBT, it only depends on the a domain outside of the light ray. The GBT for the optical geometry of the black hole is defined as follows \cite{gibbons1}:
\begin{equation}
\int\int_D K \rm d S + \int_{\partial D} \kappa \rm d t +\sum_i \alpha_i = 2\pi \chi(D),
\label{gb}
\end{equation}
where $\chi$ is the Euler characteristic, $\kappa$ is a geodesics curvature and $\alpha_i$ are the exterior angles. GW showed that the one can calculate the deflection angle in weak field limits using the GBT in this simple form \cite{gibbons1}:

\begin{equation}
\hat{\alpha}=-\int \int_{D_\infty} K  \mathrm{d}S.
\end{equation}

To solve the integral and find the deflection angle in leading-order terms, the infinite region of the surface $D_\infty$ which is bounded by photons and the straight line approximation of the photon are used. Afterwards, Werner extended this method for the Kerr black holes using the osculating Riemannian metric \cite{werner}. It is important to note that the method of GW is only working for asymptotically flat observers and sources, however Ishihara et al.
showed that it is possible to use the finite distances \cite{asahi1}.

Moreover, one of the aims of the current paper is to study the effect of phantom scalar field (phantom dilaton) and of the phantom black hole charge on gravitational lensing and in particular to the angle of deflection of the light rays. Calculating the light deflection angle caused by black holes in the weak deflection limit we study the possible manifestation of dark energy at a distances much larger than the black hole gravitational radius. For this purpose we model DE with phantom dilaton and compare the calculated light deflection angle in the geometries of the standard Einstein-Maxwell black hole and Einstein-Maxwell-anti-dilaton (EMaD) black hole which has a phantom dilaton applying the GBT to the optical geometry.


It has been recently shown that, by using the optical geometry, we can calculate the Gaussian optical curvature $\mathcal{K}$ to find the asymptotic bending angle which, in the case of asymptotically flat spacetimes, can be calculated as follows:
\begin{equation}
\hat{\alpha}=-\int \int_{D_\infty} \mathcal{K}  \mathrm{d}S,
\end{equation}
which gives exact results for bending angle. It is quite remarkable that one can compute the deflection angle by integrating over a domain outside the impact parameter.

In this paper our main aim is to obtain the deflection angle by phantom black holes \cite{Clement:2009ai,AzregAinou:2011rj}, and phantom wormholes \cite{lobo} in the weak gravitational field approximation using the GBT.

This paper is organized as follows. In section II we briefly review the phantom black holes such as GHS Black Hole and EMaD Black Hole. In section III, we calculate the deflection angle by GHS Black Hole using the GBT in the weak deflection limit. In section IV, we calculate the weak deflection angle for the EMaD Black Hole. In Section V, we shall calculate the deflection angle by three particular wormhole solutions. In Section VI we verify our findings with the help of geodesics approach. Finally, in Section VII, we comment on our results.

\section{Phantom Black Holes}

The black-hole solutions that we have studied have been obtained in the frame of the Einstein--(anti--) Maxwell dilaton theory. When phantom dilaton and/or phantom electromagnetic field is considered the action of Einstein-Maxwell-dilaton theory is generalized to the following form
\begin{equation}\label{action1}
S=\int dx^{4}\sqrt{-g}\left[  R - 2\eta_{1}g^{\mu\nu}\nabla_{\mu}\varphi\nabla_{\nu }\varphi + e^{-2\alpha\varphi}F^{\mu\nu}F_{\mu\nu}\right].  \nonumber
\end{equation}
$R$ denotes the Ricci scalar curvature, $\varphi$ is the dilaton, $F$ is the Maxwell tensor and the constant $\alpha$ determines the coupling between the dilaton and the electromagnetic field. For the usual dilaton the dilaton-gravity coupling constant $\eta_1$ takes the value $\eta_{1}=1$ while for phantom dilaton $\eta_{1}=-1$.

\subsection{Garfinkle-Horowitz-Str\"{o}minger dilaton solution}

 The line element of the static, spherically-symetric Garfinkle-Horowitz-Str\"{o}minger black hole solution of the Einstein Maxwell scalar field equations obtained
in low energy string theory is given \cite{Garfinkle:1990qj,Clement:2009ai} by

\be
ds^2 = -F(r)dt^2 + F(r)^{-1}dr^2+H(r)(d\theta^2 +
\sin^2\theta d \phi^2),\label{GMHS}
\ee
with functions $F(r)=\left(1-{r_+\over r}\right) \left(1-{r_-\over
r}\right)^{\gamma}$ and $H(r)=r^2 \left(1-{r_-\over r}\right)^{1-\gamma}$, where the parameter $\gamma=(1-\alpha^2)/(1+\alpha^2)$ has been introduced for convenience. It varies in the interval $[-1,1]$ for $\alpha\in(-\infty,\infty)$, so the stronger coupling corresponds to lower values of $\gamma$. The coresponidng solutions for the dilaton and the Maxwell field are
\be
e^{2\alpha\varphi}=\left(1-\frac{r_{-}}{r}\right)^{1-\gamma},  \quad\quad F=\frac{Q}{r^2}dt\wedge dr.
\ee
It is noted that the ADM mass $M$ and the charge $Q$ in terms of the two independent metric parameters $r_-$ and $r_+$ are
\be
2M=r_{+}+\gamma r_-, \quad\quad 2Q^2=(1+\gamma)r_+r_-, \label{charge_mass}
\ee
whence the parameters could be expressed respectively as follows

\begin{eqnarray}
r_+&=&M\left[1+\sqrt{1-{2\gamma \over 1+\gamma}\left({Q\over M}\right)^2}\right],\\
r_-&=&{M\over\gamma}\left[1-\sqrt{1-{2\gamma \over 1+\gamma}\left({Q\over M}\right)^2}\right].
\end{eqnarray}
The metric functions $F(r)$ and $H(r)$ in term of the ADM mass $M$ and the charge $Q$ than respectively are
\begin{eqnarray}\notag
 F(r)&=&\left[1-{\frac {M}{r} \left( 1+\sqrt {1-2\,{\frac {{Q}^{2}\gamma}{
 \left( 1+\gamma \right) {M}^{2}}}} \right) } \right]\\
 &\times& \left[ 1-{
\frac {M}{r\gamma } \left( 1-\sqrt {1-2\,{\frac {{Q}^{2}\gamma}{ \left( 1+
\gamma \right) {M}^{2}}}} \right) } \right] ^{\gamma},
\end{eqnarray}
\be
H(r)={r}^{2} \left[1-{\frac {M}{r\gamma} \left( 1-\sqrt {1-2\,{\frac {{Q}^{2}
\gamma}{ \left( 1+\gamma \right) {M}^{2}}}} \right) } \right] ^{1-
\gamma}.
\ee

The GHS static black hole solution (\ref{GMHS}) is characterized with an event horizon with spherical
topology, which is the biggest root of the equation $F(r)=0$ and is given by $r_{+}$. For $-1<\gamma<1$ the metric parameter $r_{-}$ represents a curvature singularity hidden inside the black hole horizon $r_{+}$. In order to see that the solution is singular at $r_{-}$ we calculate the Ricci scalar curvature and find
\begin{equation}\label{R_EMD}
    R=\frac{(1-\gamma)(1+\gamma)r_{-}^2}{2r^4}\left(1-\frac{r_{-}}{r}\right)^{\gamma-2}\left(1-\frac{r_{+}}{r}\right)
\end{equation}

The two black hole parameters $r_{-}$ and $r_{+}$ merge at
\be
\left(\frac{Q}{M}\right)^{2}=\left(\frac{Q}{M}\right)_{\rm crit}^{2}=\frac{2}{1+\gamma},
\ee
and the solution reduces to the founded by I.Z. Fisher in 1948 \cite{Fisher} massless scalar field solution, describing a naked singularity \footnote{Notice also the phantom counterpart of the massless scalar field solution proposed by O. Bergman and R. Leipnik, which have considered the spacetime of a static and sphericaly symmetric scalar field \cite{Bergmann}.}. In this case, at $r_{-}= 2M/\gamma$ a singularity is reached and
 $\gamma\in[0, 1]$. In the particular case $\gamma = 0$, the solution coincides with static, spherically
symmetric Gibbons-Maeda-Garfinkle-Horovitz-Str\"{o}minger (GMGHS) charged solution investigated in weak deflection limit regime from Keeton and Petters in \cite{KeetonPetters}. In the limit case $\gamma\rightarrow 1$ the solution restores the Reissner-Nordstrom black hole. In the particular case of absence of electrical charge $Q=0$, the Schwarzschild black hole is recovered with an event horizon $r_{+}=2M$. In this work we will restric our considerations to weak gravitational lening of black holes.

\subsection{Einstein Maxwell anti-dilaton solution}

The line element of the Einstein Maxwell anti-dilaton black hole is given by \cite{Clement:2009ai}

\be
ds^2 = -F(r)dt^2 + F(r)^{-1}dr^2+H(r)(d\theta^2 +
\sin^2\theta d \phi^2),\label{EMaDsol}
\ee
with functions $F(r)=\left(1-{r_+\over r}\right) \left(1-{r_-\over
r}\right)^{1/\gamma}$ and $H(r)=r^2 \left(1-{r_-\over r}\right)^{1-1/\gamma}$. The solutions for the dilaton and the Maxwell field are
\be
e^{2\alpha\varphi}=\left(1-\frac{r_{-}}{r}\right)^{1-1/\gamma},  \quad\quad F=\frac{Q}{r^2}dt\wedge dr.
\ee

The ADM mass $M$ and the charge $Q$  can be expressed by $r_{+}$ and $r_{-}$ in the following way
\be
2M=r_{+}+\frac{1}{\gamma}r_{-},     \quad\quad   2Q^{2}=\frac{(1+\gamma)}{\gamma}r_{+}r_{-},
\ee
whence the black hole parameters represented in therms of the ADM mass $M$ and the charge $Q$ are
\be
2M=r_++\gamma r_-, \quad\quad 2Q^2=(1+\gamma)r_+r_-, \label{charge_mass}
\ee
wherefrom the black hole parameters could be expressed respectively as follows

\begin{eqnarray}
r_+&=&M\left[1+\sqrt{1-{2\over 1+\gamma}\left({Q\over M}\right)^2}\right],\\
r_-&=&{\gamma M}\left[1-\sqrt{1-{2 \over 1+\gamma}\left({Q\over M}\right)^2}\right].
\end{eqnarray}

One can show the metric function $F(r)$ and $H(r)$ in terms of the ADM mass $M$ and the charge $Q$ respectively are
\begin{eqnarray}\notag
	F(r)&=&\left[1-\frac{M}{r}\left(1+\sqrt{1-2{Q^2\over(1+\gamma)M^2}}\right)\right]\\
    &\times &\left[1-\frac{M\gamma}{r}\left(1-\sqrt{1-2\frac{Q^2}{(1+\gamma)M^2}}\right)\right]^{1/\gamma},
\end{eqnarray}

\begin{equation}
H(r)={r}^{2} \left[1-{\frac {M\gamma}{r} \left( 1-\sqrt {1-2\,{\frac {{Q}^{2}
\gamma}{ \left( 1+\gamma \right) {M}^{2}}}} \right) } \right] ^{1-1/\gamma}.
\end{equation}

The parameter $r_{+}$ in the EMaD static black hole solution (\ref{EMaDsol}) represents the black hole horizon, which is the biggest root of the equation $F(r)=0$. For $-1<\gamma<1$ the metric parameter $r_{-}$ give rise to a curvature singularity, which can be seen from the Ricci scalar calculated for the given metric
\begin{equation}\label{R_EMaD}
    R=\frac{(\gamma-1)(1+\gamma)r_{-}^2}{2\gamma^2r^4}\left(1-\frac{r_{-}}{r}\right)^{\frac{1}{\gamma}-2}\left(1-\frac{r_{+}}{r}\right)
\end{equation}

The condition for both parameters $r_{-}$ and $r_{+}$ to be real is
\be
\left(\frac{Q}{M}\right)^{2}\leq\left(\frac{Q}{M}\right)_{\rm crit}^{2}=\frac{1+\gamma}{2}.
\ee
In the limit case $\gamma\rightarrow 1$ the Reissner-Nordstr\"{o}m black hole is restored. For $r_{-}= 0$, when the black hole charge $Q$ vanish or $\gamma\rightarrow 0$ the Schwarzschild black hole is restored.

\section{Gauss-Bonnet Theorem and Weak Gravitational Lensing by GHS and EMAD Black Holes}

\subsection{Deflection angle by Garfinkle-Horowitz-Str\"{o}minger dilaton black hole}

In this subsection we study the weak gravitational lensing in the geometry of the GHS black holes with the GBT. First we use the null geodesic $\mathrm{d}s^{2}=0$ where with the deflection angle of light in the equatorial plane $\theta =\pi /2$, we obtain the optical metric of GHS BH as follows:
\begin{equation}
\mathrm{d}t^{2}=\frac{1}{F(r)^2}\mathrm{d}r^2+\frac{H(r)}{F(r)}\mathrm{d}\varphi^2.
\end{equation}

Afterwards, we make the transformation to tortoise coordinates with  $r^{\star }$, and find the
$f(r^{\star })$
\begin{equation}
\mathrm{d}r^{\star }=\frac{1}{F(r)}\mathrm{d}r,\,\,\,f^2(r^{\star })=\frac{H(r)}{F(r)}.
\end{equation}

The metric of GMGHS reduces to
\begin{equation}
\mathrm{d}t^{2}=\tilde{g}_{ab}\,\mathrm{d}x^{a}\mathrm{d}x^{b}=\mathrm{d}{%
r^{\star }}^{2}+f^{2}(r^{\star })\mathrm{d}\varphi ^{2}. \,\,\,\,(a,b=r,\varphi),
\end{equation}

Then we obtain the Gaussian optical curvature $K$:
\begin{eqnarray}
  \mathcal{K} & = & - \frac{1}{f (r^{\star})}  \frac{\mathrm{d}^2 f
  (r^{\star})}{\mathrm{d} r^{\star 2}} \\\notag
  & = & - \frac{1}{f (r^{\star})}  \left[ \frac{\mathrm{d} r}{\mathrm{d}
  r^{\star}}  \frac{\mathrm{d}}{\mathrm{d} r} \left( \frac{\mathrm{d}
  r}{\mathrm{d} r^{\star}} \right) \frac{\mathrm{d} f}{\mathrm{d} r} + \left(
  \frac{\mathrm{d} r}{\mathrm{d} r^{\star}} \right)^2 \frac{\mathrm{d}^2
  f}{\mathrm{d} r^2} \right] .
\end{eqnarray}

Since we are interested in the weak limit, we can approximate the optical Gaussian curvature as

\begin{align}\label{Curvature1}
\begin{split}
\mathcal{K} \approx & -{\frac {\gamma\,{\it r_-}}{{r}^{3}}}-{\frac {{\it r_+}}{{r}^{3}}}+\,
{\frac {{3{\it r_+}}^{2}}{4{r}^{4}}}+\,{\frac {9\gamma\,{\it r_-}\,{\it
r_+}}{2{r}^{4}}} \\ &+\,{\frac {{{\it r_-}}^{2} \left( 8\,{\gamma}^{2}-6\,
\gamma+1 \right) }{4{r}^{4}}}
\end{split}
\end{align}

Now, we calculate the deflection angle using the Gaussian optical curvature. For this purpose, we select a non-singular region $\mathcal{D}_{R}$ with boundary $\partial
\mathcal{D}_{R}=\gamma _{\tilde{g}}\cup C_{R}$. Note that this region allows the GBT to be stated as follows:
\begin{equation}
\iint\limits_{\mathcal{D}_{R}}\mathcal{K}\,\mathrm{d}S+\oint\limits_{\partial \mathcal{%
D}_{R}}\kappa \,\mathrm{d}t+\sum_{i}\theta _{i}=2\pi \chi (\mathcal{D}_{R}),
\end{equation}
in which $\kappa $ gives the geodesic curvature, $\mathcal{K}$ stands for the Gaussian optical curvature, while $\theta_{i}$ is the exterior angle at the $i^{th}$ vertex. We can choose a non-singular domain outside of the light ray with the Euler characteristic number $%
\chi (\mathcal{D}_{R})=1$. In order to find the deflection angle, let us first compute the geodesic curvature using the following relation
\begin{equation}
\kappa =\tilde{g}\,\left(\nabla _{\dot{%
\gamma}}\dot{\gamma},\ddot{\gamma}\right)
\end{equation}
together with the unit speed condition $\tilde{g}(\dot{\gamma},\dot{%
\gamma})=1$, where $\ddot{\gamma}$ gives the unit acceleration vector. If we let $R\rightarrow \infty $, our two jump angles ($\theta _{\mathcal{O}}$, $%
\theta _{\mathcal{S}}$) become $\pi /2,$ or in other words, the sum of jump angles to the source $\mathcal{S}$, and observer $\mathcal{O}$, satisfies $\theta _{\mathit{O}%
}+\theta _{\mathit{S}}\rightarrow \pi $. Hence we can write GBT as
\begin{equation}
\iint\limits_{\mathcal{D}_{R}}\mathcal{K}\,\mathrm{d}S+\oint\limits_{C_{R}}\kappa \,%
\mathrm{d}t\overset{{R\rightarrow \infty }}{=}\iint\limits_{\mathcal{D}%
_{\infty }}\mathcal{K}\,\mathrm{d}S+\int\limits_{0}^{\pi +\hat{\alpha}}\mathrm{d}\varphi
=\pi.
\end{equation}

Let us now compute the geodesic curvature $\kappa$. To do so, we first point out that $\kappa (\gamma _{\tilde{g}})=0$, since $\gamma _{\tilde{g}}$ is a
geodesic. We are left with the following
\begin{equation}
\kappa (C_{R})=|\nabla _{\dot{C}_{R}}\dot{C}_{R}|,
\end{equation}
where we choose $C_{R}:=r(\varphi)=R=\text{const}$. The radial part is evaluated as
\begin{equation}
\left( \nabla _{\dot{C}_{R}}\dot{C}_{R}\right) ^{r}=\dot{C}_{R}^{\varphi
}\,\left( \partial _{\varphi }\dot{C}_{R}^{r}\right) +\tilde{\Gamma} _{\varphi
\varphi }^{r}\left( \dot{C}_{R}^{\varphi }\right) ^{2}. \label{12}
\end{equation}

From the last equation, it  obvious that the first term vanishes, while the second term is calculated using Eq. (7) and the unit speed condition. For the geodesic curvature we find
\begin{eqnarray}\notag
\lim_{R\rightarrow \infty }\kappa (C_{R}) &=&\lim_{R\rightarrow \infty
}\left\vert \nabla _{\dot{C}_{R}}\dot{C}_{R}\right\vert , \notag \\
&\rightarrow &\frac{1}{R}.
\end{eqnarray}%

On the other hand, for very large radial distance yields
\begin{eqnarray}
\lim_{R\rightarrow \infty } \mathrm{d}t&\to & \left(R-\frac{(\gamma-1)r_-}{2 \gamma}\right) \, \mathrm{d}\varphi.
\end{eqnarray}%

If we combine the last two equations, we find $
\kappa (C_{R})\mathrm{d}t= \mathrm{d}\,\varphi
$. It is convenient to choose the deflection line as $r=b/\sin \varphi$, in that case, the deflection angle from Eq. (30) can be recast in the following from
\begin{eqnarray}\label{int0}
\hat{\alpha}=-\int\limits_{0}^{\pi}\int\limits_{\frac{b}{\sin \varphi}}^{\infty}\mathcal{K}\mathrm{d}S.
\end{eqnarray}

If we substitute Eq. \eqref{Curvature1} into the last equation and solve it. Note that we use the following relation $\mathrm{d}r^{\star} \approx \mathrm{d}r$, valid in the limit as $R\to \infty $. One can easily solve this integral up to the second order terms to find the following result
\begin{equation}\label{GB10}
\begin{split}
\hat{\alpha}&\simeq 2\,{\frac {{\it r_+}}{b}}+ 2\,{\frac {\gamma\,{\it r_-}}{b}}-\frac{3\pi}{16}\frac{r_{+}^2}{b^2}\\&-\frac{\pi(1 - 6\gamma + 8\gamma^2)r_{-}^2}{16b^2}-\frac{9\pi \gamma}{8}\frac{r_{+}r_{-}}{b^2}.
\end{split}
\end{equation}

\subsection{Deflection angle by  Einstein Maxwell anti-dilaton black hole}

In the current subsection we consider the deflection angle of EMaD black hole in weak field approximation using the GBT. First we use the null geodesic $\mathrm{d}s^{2}=0$ where with the deflection angle of light in the equatorial plane $\theta =\pi /2$, we obtain the optical metric of EMaD black hole as follows:
\begin{equation}
\mathrm{d}t^{2}=\frac{1}{F(r)^2}\mathrm{d}r^2+\frac{H(r)}{F(r)}\mathrm{d}\varphi^2,
\end{equation}
with functions $F(r)=\left(1-{r_+\over r}\right) \left(1-{r_-\over
r}\right)^{1/\gamma}$ and $H(r)=r^2 \left(1-{r_-\over r}\right)^{1-1/\gamma}$.

For this purpose, similarly to previous section, first we obtain the Gaussian optical curvature $\mathcal{K}$. Since we are interested in the weak limit, we can approximate the optical Gaussian curvature as

\begin{align}\label{Curvature2}
\begin{split}
\mathcal{K} \approx & -{\frac {{\it r_+}}{{r}^{3}}}-{\frac {{\it r_-}}{{r}^{3}\gamma}}+\,{
\frac {{3{\it r_+}}^{2}}{4{r}^{4}}}+\,{\frac {9{\it r_-}\,{\it r_+}}{2{r}^
{4}\gamma}} \\ & + \,{\frac { \left( {\gamma}^{2}-6\,\gamma+8 \right) {{
\it r_-}}^{2}}{4{r}^{4}{\gamma}^{2}}}.
\end{split}
\end{align}

For the geodesic curvature we find
\begin{eqnarray}\notag
\lim_{R\rightarrow \infty }\kappa (C_{R}) &=&\lim_{R\rightarrow \infty
}\left\vert \nabla _{\dot{C}_{R}}\dot{C}_{R}\right\vert , \notag \\
&\rightarrow &\frac{1}{R}.
\end{eqnarray}%

On the other hand, for very large radial distance yields
\begin{eqnarray}
\lim_{R\rightarrow \infty } \mathrm{d}t&\to & \left(R-\frac{(\gamma-1)r_-}{2 \gamma}\right) \, \mathrm{d}\varphi.
\end{eqnarray}%

If we combine the last two equations, we find $
\kappa (C_{R})\mathrm{d}t= \mathrm{d}\,\varphi
$. It is convenient to choose the deflection line as $r=b/\sin \varphi$, in that case, the deflection angle from Eq. (30) can be recast in the following from
\begin{eqnarray}\label{int0}
\hat{\alpha}=-\int\limits_{0}^{\pi}\int\limits_{\frac{b}{\sin \varphi}}^{\infty}\mathcal{K}\mathrm{d}S.
\end{eqnarray}

If we substitute Eq. \eqref{Curvature2} into the last equation, and solve the integral in the leading order terms, we obtain the following result for deflection angle:
\begin{equation}\label{GB1}
\begin{split}
\hat{\alpha}&\simeq 2\,{\frac {{\it r_+}}{b}}+ 2\,{\frac {\,{\it r_-}}{b\gamma}}-\frac{3\pi}{16}\frac{r_{+}^2}{b^2}\\&-\frac{\pi(8 - 6\gamma + \gamma^2)r_{-}^2}{16\gamma^4 b^2}-\frac{9\pi}{8\gamma}\frac{r_{+}r_{-}}{b^2}.
\end{split}
\end{equation}

\section{Light deflection by Phantom Wormholes}
\subsection{Wormholes with a bounded mass function}
Recently a class of wormhole solutions with a phantom energy equation of state parameter $\omega<-1$ was reported in Ref. \cite{lobo}. As a first example we shall examine a particular wormhole solution with a bounded mass function with the metric written as follows \cite{lobo}
\begin{equation}\label{WSBMF}
ds^2=-\left(1+\frac{a r_0}{r}\right)^{1-\frac{1}{a}}dt^2+\frac{dr^2}{1-\frac{r_0}{r}\left(\frac{a r_0}{r}+1-a   \right)}+r^2 d\Omega^2,
\end{equation}
where the parameter $a$ belongs to the interval $-1<a<0$. Note that the wormhole throat which connects
two asymptotic regions is located at $r_0$, with the following condition $b(r_0)=r_0$. The mass function with a finite value is given by
\begin{equation}
m(r)=\frac{a r_0}{2}\left(\frac{r_0}{r}-1   \right).
\end{equation}

Let is write the wormhole optical metric in the $(r,\varphi)$ plane which takes the form
\begin{equation}
dt^2=\frac{dr^2}{\left(1+\frac{a r_0}{r}\right)^{1-\frac{1}{a}}\left[1-\frac{r_0}{r}\left(\frac{a r_0}{r}+1-a   \right)\right]}+\frac{r^2 d\varphi^2}{\left(1+\frac{a r_0}{r}\right)^{1-\frac{1}{a}}}.
\end{equation}

The Gaussian optical curvature on the other hand is approximated as follows
\begin{equation}
\mathcal{K} \simeq \frac{\left(7-14a\right)r_0^2+4r\left(a-1\right)r_0}{4 r^4}.
\end{equation}

In addition the geodesic curvature reads
\begin{eqnarray}\notag
\lim_{R\rightarrow \infty }\kappa (C_{R}) &=&\lim_{R\rightarrow \infty
}\left\vert \nabla _{\dot{C}_{R}}\dot{C}_{R}\right\vert , \notag \\
&\rightarrow &\frac{1}{R}.
\end{eqnarray}%

On the other hand, for very large radial distance we have
\begin{eqnarray}
\lim_{R\rightarrow \infty } \mathrm{d}t&\to & R \mathrm{d}\varphi.
\end{eqnarray}%

To put it another way, our optical-wormhole metric is asymptotically Euclidean satisfying the condition $\kappa (C_{R})\mathrm{d}t/\mathrm{d}\varphi =1$. Moreover if we substitute  the expression for the Gaussian optical curvature in Eq. (26) the deflection angle is recast as follows
\begin{eqnarray}\label{int0}
\hat{\alpha}=-\int\limits_{0}^{\pi}\int\limits_{\frac{b}{\sin \varphi}}^{\infty}\left( \frac{\left(7-14a\right)r_0^2+4r\left(a-1\right)r_0}{4 r^3}   \right)\mathrm{d}r d\varphi.
\end{eqnarray}

As a result, up to the first oder approximation we find the following result for the deflection angle:
\begin{equation}\label{WSDAGBT}
\hat{\alpha}\simeq \frac{2\,r_0}{b}\left(1-a   \right)-\frac{ \pi r_0^2}{16 b^2}\left(7-14 a \right).
\end{equation}
The obtained result (\ref{WSDAGBT}) coinsides up to the first order with the approximate expression for the weak deflection angle of the light calculated in Schwarzschild spacetime, in the limit $a\rightarrow 0$, where because of that the wormhole is asymptotically flat $r_{0}=2M$. Here the AMD mass $M$ is involved.

\subsection{Wormholes with an unbounded mass function}
Our second example is a wormhole solution with unbounded mass function with the following spacetime metric \cite{lobo}
\begin{equation}\label{WSUMF}
ds^2=-\left(1+\frac{1-a}{\sqrt{r/r_0}}\right)^{2}dt^2+\frac{dr^2}{1-\frac{a}{\sqrt{r/r_0}}-\frac{1-a}{r/r_0}}+r^2 d\Omega^2 .
\end{equation}

With the mass function given as
\begin{equation}
m(r)=\frac{a}{2}\left(  \sqrt{r r_0}-r_0\right).
\end{equation}

It is noted that, the above expression is positive throughout the spacetime, but unbounded as $r \to \infty $.  Furthermore the parameter $a$ lies in the range $0<a<2$.
The corresponding optical metric in the $(r,\varphi)$ reads
\begin{equation}
dt^2=\frac{dr^2}{\left(1+\frac{1-a}{\sqrt{r/r_0}}\right)^{2}\left(1-\frac{a}{\sqrt{r/r_0}}-\frac{1-a}{r/r_0}\right)}+\frac{r^2 d\varphi^2}{\left(1+\frac{1-a}{\sqrt{r/r_0}}\right)^{2}},
\end{equation}
with the Gaussian optical curvature
\begin{equation}
\mathcal{K} \simeq - \frac{r_0}{2 r^3}+\frac{\sqrt{r_0}}{4 r^{5/2}}-\frac{3 r_0^{3/2}}{2 r^{7/2}}-\frac{3 r_0^2}{4 r^4}+\Xi(r,r_0)\, a
\end{equation}
in which
\begin{equation}
\Xi(r,r_0)=- \frac{3 r_0}{8 r^3}-\frac{\sqrt{r_0}}{2 r^{5/2}}+\frac{21 r_0^{3/2}}{8 r^{7/2}}+\frac{9 r_0^2}{4 r^4}.
\end{equation}

The geodesic curvature reveals that
\begin{eqnarray}\notag
\lim_{R\rightarrow \infty }\kappa (C_{R}) &=&\lim_{R\rightarrow \infty
}\left\vert \nabla _{\dot{C}_{R}}\dot{C}_{R}\right\vert , \notag \\
&\rightarrow &\frac{1}{R},
\end{eqnarray}%
together with
\begin{eqnarray}
\lim_{R\rightarrow \infty } \mathrm{d}t&\to & R \mathrm{d}\varphi.
\end{eqnarray}%

From these results it is possible to show $
\kappa (C_{R})\mathrm{d}t= \mathrm{d}\,\varphi
$. It is a straightforward analyses to check that the light ray equation is governed by equation $r=b/\sin \varphi$. The GBT implies
\begin{equation}
\hat{\alpha}=-\int\limits_{0}^{\pi}\int\limits_{\frac{b}{\sin \varphi}}^{\infty}\left( - \frac{r_0}{2 r^3}+\frac{\sqrt{r_0}}{4 r^{5/2}}-\frac{3 r_0^{3/2}}{2 r^{7/2}}-\frac{3 r_0^2}{4 r^4}+\Xi\, a \right)r\mathrm{d}r d\varphi.
\end{equation}

Solving the integral we obtain the following expression
\begin{eqnarray}\label{WSDA2}
\hat{\alpha}&\simeq & - \frac{\sqrt{\pi} \sqrt{r_0} \,\Gamma\left(\frac{3}{4}\right)}{2 \sqrt{b}\,\Gamma\left(\frac{5}{4}\right)}(1-2a) + \frac{r_0}{4b}(4+3a) \\\notag
&+& \frac{\sqrt{\pi}r_0^{3/2} \Gamma\left(\frac{5}{4}\right)}{6 b^{3/2}\,\Gamma\left(\frac{7}{4}\right)}(6-\frac{21}{2}a) + \frac{\pi r_0^2}{16 b^2}(3-9 a).
\end{eqnarray}

\subsection{Wormholes with vanishing redshift function}
Our last example, is a rather simple wormhole solution with a vanishing redshift function. The spacetime metric in that case is given by \cite{lobo}
\begin{equation}\label{WSVRF}
ds^2=-dt^2+\frac{dr^2}{1-(r_0/r)^{1-\alpha}}+r^2 d\theta^2+r^2 \sin^2\theta d\varphi^2,
\end{equation}
in which $0<\alpha<1$. The optical metric takes the form
\begin{equation}
dt^2=\frac{dr^2}{1-(r_0/r)^{1-\alpha}}+r^2  d\varphi^2,
\end{equation}
and
\begin{equation}
\mathcal{K} = - \frac{(1-\alpha) r_0}{2 r^3}\left(\frac{r_0}{r}  \right)^{-\alpha}.
\end{equation}


It is a straightforward to see that
\begin{eqnarray}\notag
\lim_{R\rightarrow \infty }\kappa (C_{R}) &=&\lim_{R\rightarrow \infty
}\left\vert \nabla _{\dot{C}_{R}}\dot{C}_{R}\right\vert , \notag \\
&\rightarrow &\frac{1}{R},
\end{eqnarray}%
thus the optical geometry is asymptotically Euclidean
\begin{eqnarray}
\lim_{R\rightarrow \infty } \frac{\kappa (C_{R})\mathrm{d}t}{\mathrm{d}\,\varphi}=1.
\end{eqnarray}

Utilizing the GBT it follows
\begin{eqnarray}
\hat{\alpha} =-\int\limits_{0}^{\pi}\int\limits_{\frac{b}{\sin \varphi}}^{\infty}\left[- \frac{(1-\alpha) r_0}{2 r^3}\left(\frac{r_0}{r}  \right)^{-\alpha}\right]r\mathrm{d}r d\varphi.
\end{eqnarray}

Finally evaluating the above integral we find
\begin{eqnarray}
\hat{\alpha}=\frac{\sqrt{\pi}}{2}\frac{\Gamma(1-\frac{\alpha}{2})}{\Gamma(\frac{3}{2}-\frac{\alpha}{2})}\left(\frac{r_{0}}{b}\right)^{1-\alpha}.
\end{eqnarray}

To simplify the problem we perform a power series expanssion of the deflection angle, which is given by
\begin{align}\label{WSVRFDA}
\hat{\alpha} & \simeq \frac{(1+\alpha[1-\ln{2}])r_{0}}{b} \\ & -\frac{(\pi^2-12[2+(\ln{2})^2-\ln{4}])r_{0}\alpha^2}{24b} +\mathcal{O}(\ln{r_0},\alpha^2). \nonumber
\end{align}

In the weak field regime the approximate expression (\ref{WSVRFDA}) for the deflection angle coincides with the light bending angle in Schwarzschild spacetime in the limit $\alpha \rightarrow 0$.

\section{Geodesics}

In this section we apply the standard geodesics approach to show that one can obtain the same results \eqref{GB10} and \eqref{GB1} for the deflection angles in the weak field regime. The geodesic equations can be derived using the Euler-Lagrange formalism \cite{Rindler} with the Lagrangian
\begin{equation}\label{Lagrangian}
    2\mathcal{L}=-F(r)\dot{t}^{\,2}+F(r)^{-1}\dot{r}^{\,2}+H(r)\dot{\phi}^{\,2}
\end{equation}
for geodesics in the $\theta=\pi/2$ hypersurface of the considered black hole or wormhole geometry. Here, the dot represents the derivative with respect to the affine parameter $\lambda$, i.e. $\dot{q}=dq/d\lambda$. Since the metric coefficients do not depend neither on $\phi$ nor $t$, than the geodesic equations corresponding to these "cyclic" coordinates will be associated to two integrals of motion according to the equations $\dot{\Pi}_{q}-\partial\mathcal{L}/\partial q=0$ leading to
\begin{equation}
    \dot{\Pi}_{t}=0, \quad\quad \dot{\Pi}_{\phi}=0.
\end{equation}
Here, $\Pi_{q}=\partial\mathcal{L}/\partial \dot{q}$ are the conjugate momenta to the spacetime coordinates $q$ and for the Lagrangian (\ref{Lagrangian}) are given by
\begin{eqnarray}\label{ConjugateMomenta}
    && \Pi_{t}=-F(r)\dot{t}\equiv-E,  \\
    && \Pi_{r}=F(r)^{-1}\dot{r} \\
    && \Pi_{\phi}=H(r)\dot{\phi}\equiv L,
\end{eqnarray}
where $E$ and $L$ are the first integrals of motion for a test particle, corresponding to the energy at infinity and the angular momentum, respectively. Therefore, the Hamiltonian for a test particle with rest mass $m$ in the static and spherically symmetric spacetimes under consideration is given by
\begin{equation}\label{HamiltonianDef}
    \mathcal{H}=\Pi_{t}\dot{t}+\Pi_{\phi}\dot{\phi}+\Pi_{r}\dot{r}-\mathcal{L},
\end{equation}
\begin{equation}\label{HamiltonianMetric}
    2\mathcal{H}=-E\dot{t}+L\dot{\phi}+F(r)^{-1}\dot{r}^2=-m^2.
\end{equation}
Considering only massless particles and taking into account the condition $m=0$, for photons we obtain the following equations of motion
\begin{eqnarray}
    && \dot{t}=\frac{E}{F(r)}, \label{EqsOfMotionT} \\
    && \dot{\phi}=\frac{L}{H(r)}, \label{EqsOfMotionPhi} \\
    && \dot{r}^2=E^{2}-L^{2}\frac{F(r)}{H(r)}. \label{EqsOfMotionR}
\end{eqnarray}

Finding the set of values of the integrals of motion $E$ and $L$, for which $\dot{r}=0$, we can determine a precise relation between the light ray impact parameter $b=L/E$ and the closest approach distance $r_{0}$
\begin{equation}\label{ImpactParameter}
    b(r_{0})=\sqrt{\frac{H(r_{0})}{F(r_{0})}}.
\end{equation}

Having the expressions (\ref{EqsOfMotionPhi}) and (\ref{EqsOfMotionR}) we find the azimuthal shift of the photon as a function of the coordinate distance
\begin{equation}\label{DphiDr}
    \frac{d\phi}{dr}=\pm\frac{1}{H(r)}\left[\frac{1}{b^2}-\frac{F(r)}{H(r)}\right]^{-1/2},
\end{equation}
where the sign in front of square root is negative along the motion of the photon from the source to the minimum distance $r_{0}$ and the sign is positive for the motion of the photon to the source, afterwards.

Let's assume the light source and observer are placed in the asymptotically flat region of the spacetime. Then, according to the symmetry in the phase of approach and departure we can write the whole light deflection angle via integration of Eq. (\ref{DphiDr}) from $r_{0}$ to infinity
\begin{equation}\label{DeflectionAngle}
    \hat{\alpha}(r_{0})=2\int_{r_{0}}^{\infty}\left|\frac{d\phi}{dr}\right|dr-\pi.
\end{equation}

\subsection{Deflection angle by Garfinkle-Horowitz-Str\"{o}minger black hole}

The relation between the impact parameter $b$ and the distance of closest approach of the light ray $r_{0}$ in the spacetime of the Garfinkle-Horovitz-Strominger dilaton black hole is given by
\begin{equation}\label{ImpactParameterGHS}
    b(r_{0})=r_{0}^{1+\gamma}\sqrt{\frac{(r_{0}-r_{-})^{1-2\gamma}}{(r_{0}-r_{+})}}.
\end{equation}
We expect for small deviations of the light rays the distance of a closest approach $r_{0}$ to be of the same order as the impact parameter, under assumptions $r_{0}\gg r_{+}$ and $r_{0}\gg r_{-}$. Therefore, for the solution of Eq. (\ref{EqsOfMotionR}) we suggest that
\begin{equation}\label{r0approximation}
    r_{0}\simeq b \left\{1+\sum\limits_{i,j=1}^{2} c_{r_{+}r_{-}}\epsilon_{r_{+}}^{i}\epsilon_{r_{-}}^{j}+\mathcal{O}(\epsilon_{r}^3)\right\},
\end{equation}
where we are introducing two independent expansion parameters in terms of the black hole horizons
\begin{equation}\label{epspm}
    \epsilon_{r_{+}}=\frac{r_{+}}{b}, \quad\quad \epsilon_{r_{-}}=\frac{r_{-}}{b}.
\end{equation}
Here the coefficients $c_{r_{+}r_{-}}$ are real numbers, and the summation is over all possible combinations of epsilon powers $i, j$ up to and including second order terms. Notice, that the approximation (\ref{r0approximation}) is not longer valid in limit $\gamma\rightarrow -1$, when the both black hole horizons diverge.

Solving the equation $\dot{r}(r_{0})=0$ we find for the closest approach distance
\begin{equation}\label{ClosestDistance}
\begin{split}
    r_{0}\simeq b & \left\{1 -\frac{r_{+}}{2b}+\frac{(1-2\gamma)r_{-}}{2b}+\frac{(1-2\gamma)r_{+}r_{-}}{4b^2} \right.  \\
                          & \quad \;\, -\left. \frac{3r_{+}^2}{8b^2}+\frac{(1-4\gamma^2)r_{-}^2}{8b^2}+\mathcal{O}(\epsilon^3) \right\}.
\end{split}
\end{equation}

The azimuthal shift of the photon follows from the Eqs. (\ref{EqsOfMotionPhi}) and (\ref{EqsOfMotionR}) and is given by
\begin{equation}\label{DpfiDrGHS}
    \frac{d\phi}{dr}=\frac{(r-r_{-})^{\gamma-1}}{r^{1+\gamma}}\left[\frac{1}{b^2}-\frac{(r-r_{+})}{r^{2(1+\gamma)}(r-r_{-})^{1-2\gamma}}\right]^{-1/2}
\end{equation}

Assuming $\varepsilon_{r_{+}}=r_{+}/r_{0}$ and $\varepsilon_{r_{-}}=r_{-}/r_{0}$ are small enough quantities, we can consider a Taylor expansion series of Eq. (\ref{DpfiDrGHS}) up to and including second order terms of the new expansion parameters $\varepsilon_{r_{+}}$ and $\varepsilon_{r_{-}}$. To make integration afterwards easier, we introduce new variable $u=r_{0}/r$, in terms of which
\begin{widetext}
\begin{equation}\label{TaylorIFuncGHS}
\begin{split}
    \frac{d\phi}{du} \simeq \frac{1}{\sqrt{1-u^2}} & +\frac{1}{2}\left[\frac{1+u+u^2}{\sqrt{1-u}(1+u)^{3/2}}\right]\varepsilon_{r_{+}}-\frac{1}{2}\left[ \frac{1-u(1+u)-2\gamma}{\sqrt{1-u}(1+u)^{3/2}} \right]\varepsilon_{r_{-}} +\frac{3}{8}\left[\frac{\left(u^2+u+1\right)^2}{\sqrt{1-u} (1+u)^{5/2}}\right]\varepsilon_{r_{+}}^2 \\ &-\frac{1}{4}\left[\frac{(1-2 \gamma )-2 \gamma  u+(1-6 \gamma ) u^2-2 u^3-u^4}{ \sqrt{1-u} (1+u)^{5/2}}\right]\varepsilon_{r_{+}}\varepsilon_{r_{-}} \\ & -\frac{1}{8}\left[\frac{ 1-4 \gamma ^2-(12 \gamma -6) u-\left(8 \gamma ^2+1\right) u^2-6 u^3-3 u^4 }{\sqrt{1-u^2} (1+u)^{2}}\right]\varepsilon_{r_{-}}^2 +\mathcal{O}(\varepsilon^3).
\end{split}
\end{equation}
\end{widetext}
Calculations of the deflection angle, according to Eq. (\ref{DeflectionAngle}), show that the zero order term $\mathcal{O}(1)$ evaluated on $\pi$ is shortening. As opposed to that the first order terms proportional to $\mathcal{O}(\varepsilon_{r_{+}})$ and $\mathcal{O}(\varepsilon_{r_{-}})$ respectively, give the most significant contributions $2\varepsilon_{r_{+}}$ and $2\gamma\varepsilon_{r_{-}}$, to the light deflection angle, as can be seen from the next expression
\begin{align}\label{DAGHSro}
   \nonumber \hat{\alpha}(r_{0}) & \simeq \frac{2r_{+}}{r_{0}}+\frac{2\gamma r_{-}}{r_{0}}-\left(\frac{3\pi}{8}-1+\frac{3(2-\pi)\gamma}{2}\right)\frac{r_{+}r_{-}}{r_{0}^2} \\
    & \nonumber + \left(\frac{15\pi}{16}-1\right)\frac{r_{+}^2}{r_{0}^2}-\left(\frac{\pi}{16}-\gamma+(2-\pi )\gamma^2\right)\frac{r_{-}^2}{r_{0}^2} \\ & +\mathcal{O}(\varepsilon_{r}^3).
\end{align}
Hence, taking into account the power series \eqref{ClosestDistance}, we can represent the light deflection angle in terms of the photon integrals of motion $L$ and $E$, via the impact parameter $b$, up to and including the second orders of $\epsilon_{r}$. Therefore, the bending of light ray by GHS black hole leads to the following components of the deflection angle
\begin{equation}\label{DAGHS}
\begin{split}
    \hat{\alpha}(b) & \simeq \frac{2r_{+}}{b}+\frac{2\gamma r_{-}}{b}-\frac{3 \pi (1 - 4\gamma)r_{+}r_{-}}{8b^2}\\
    & +\frac{15\pi}{16}\frac{r_{+}^2}{b^2}-\pi\left(\frac{1}{16}-\gamma^2\right)\frac{r_{-}^2}{b^2} +\mathcal{O}(\epsilon_{r}^3).
\end{split}
\end{equation}
Up to the first formal order in the expansion parameter the approximate expression for the light deflection angle (\ref{DAGHS}) coincides with the result \eqref{GB10} found by GBT.



\subsection{Deflection angle by Einstein-Maxwell anti-dilaton black hole}

In the case of Einstein-Maxwell anti-dilaton black hole the impact parameter of the photon $b$ and the distance of closest approach $r_{0}$ obey the following relation
	\begin{equation}\label{IParameterEMaD}
    b(r_{0})=r_{0}^{1+\frac{1}{\gamma}}\sqrt{\frac{(r_{0}-r_{-})^{1-\frac{2}{\gamma}}}{(r_{0}-r_{+})}}.
\end{equation}

The azimuthal shift of the photon as a function of the coordinate distance is given by
\begin{equation}\label{DpfiDrEMaD}
	\frac{d\phi}{dr}=\frac{(r-r_{-})^{\frac{1}{\gamma}-1}}{r^{1+\frac{1}{\gamma}}}		\left[\frac{1}{b^2}-\frac{(r-r_{+})}{r^{2+\frac{2}{\gamma}}(r-r_{-})^{1-\frac{2}	 {\gamma}}}\right]^{-1/2}
\end{equation}

\begin{figure*}[t!]
            \centering
            \includegraphics[width=\textwidth]{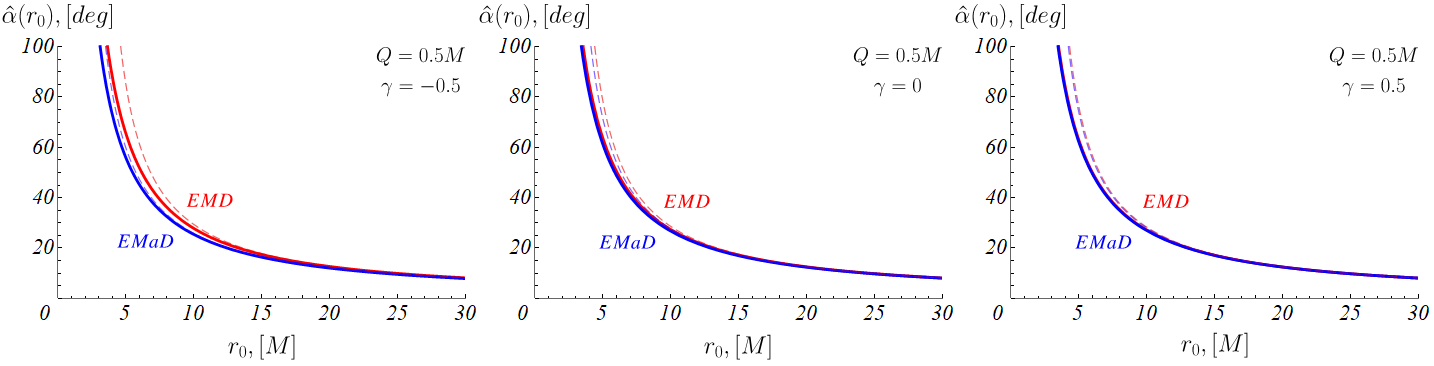} \\[1mm]	
            \caption{\label{fig:1}\small \textit{The deflection angles by GHS (red) and EMaD (blue) black holes as a function of the closest approach distance $r_{0}$ for an electrical charge $Q=0.5M$ and three values of $\gamma$: $\gamma=-0.5$ (left), $\gamma=0$ (middle), and $\gamma=-0.5$ (right). The dashed lines represent exact numerical calculations of the deflection angle, as opposed to solid lines referring to approximated analytically results \eqref{DAGHSro} and \eqref{DAEMaDro} yield by the geodesics.}} \label{fig:1}
\end{figure*}

Introducing two independent expansion parameters $\varepsilon_{r_{+}}=r_{+}/r_{0}$ and $\varepsilon_{r_{-}}=r_{-}/r_{0}$, we can perform Taylor series expansion of Eq. (\ref{DpfiDrEMaD}) up to and including second order terms. Therefore, the bending angle of the light ray, defined by Eq. (\ref{DeflectionAngle}) is given by
\begin{align}\label{DAEMaDro}
   \nonumber \hat{\alpha}(r_{0}) & \simeq  \frac{2r_{+}}{r_{0}}+\frac{2r_{-}}{\gamma b}+\frac{(3 \pi  (4-\gamma )-8 (3-\gamma ))r_{+}r_{-}}{8 \gamma r_{0}^2} \\
    & \nonumber + \left(\frac{15\pi}{16}-1\right)\frac{r_{+}^2}{r_{0}^2}-\left(\frac{\pi }{16}-\frac{\pi+\gamma-2}{\gamma ^2}\right)\frac{r_{-}^2}{r_{0}^2} \\ & + \mathcal{O}(\varepsilon_{r}^3).
\end{align}

Here, similar to the procedure described in the previous paragraph we take into account the power series \eqref{r0approximation}. After completing approximation procedure, we obtain the light deflection angle
\begin{equation}\label{DAEMaD}
\begin{split}
    \hat{\alpha}(b) & \simeq \frac{2r_{+}}{b}+\frac{2r_{-}}{\gamma b}-\frac{3 \pi}{8} \left(1-\frac{4}{\gamma}\right)\frac{r_{+}r_{-}}{b^2}\\
    & +\frac{15\pi}{16}\frac{r_{+}^2}{b^2}+\pi\left(\frac{1}{\gamma^2}-\frac{1}{16}\right)\frac{r_{-}^2}{b^2} +\mathcal{O}(\epsilon_{r}^3).
\end{split}
\end{equation}
The most significant contributions to the light deflection angle in that case are given by the first order terms $2\epsilon_{r_{+}}$ and $2\epsilon_{r_{-}}/\gamma$, proportional to $\mathcal{O}(\epsilon_{r_{+}})$ and $\mathcal{O}(\epsilon_{r_{-}})$, respectively.

As expected, up to the given formal order in the expansion parameters the deflection angle in the weak limit approximation is found to be the same result (\ref{GB1}) found by GBT.

Both deflection angles are plotted in Fig. \ref{fig:1}. The main difference in the behaviour of the deflection angles manifests in the changing of the metric parameter $\gamma$. In the case of EMD black hole, the deflection angle increases with opposed to the bending angle by EMaD black hole, which decreases with the increase of the parameter $\gamma$. The light ray suffers equal deflection by any of the black holes when $\gamma$ goes to 1, no matter what is the electrical charge. Furthermore, both deflection angles decrease with the increases of the charge. In general, the parameter $\gamma$ and the electrical charge $Q/M$ have a negligible impact on the observable quantities as can be seen from the discussion in Sec. \ref{sec:VI}.

\subsection{Deflection angle by phantom wormholes}

In the case of the wormhole solution with a bounded mass function \eqref{WSBMF}, the impact parameter of the photon $b$ and the distance of closest approach $r_{m}$ obey the following relation
\begin{equation}\label{IParameterWBMF}
 b(r_{m})=r_{m}\left(1-\frac{a r_{0}}{r_{m}}\right)^{\frac{1}{2}\left(\frac{1}{a}-1\right)}.
\end{equation}

The azimuthal shift of the photon follows from the Eqs. (\ref{EqsOfMotionPhi}) and (\ref{EqsOfMotionR}) and is given by

\begin{equation}\label{WSASHift}
	\frac{d\phi}{dr}=\frac{1}{r^2}\left[\frac{1}{b^2}-\frac{1}{r^2}\left(1-\frac{a r_{0}}{r}\right)^{1-\frac{1}{a}}\right]^{-1/2}.
\end{equation}

In order to obtain the light deflection angle in weak deflection limit, we perform a Taylor series expansion of \eqref{WSASHift} up to and including second order terms in the powers of new small expansion parameter $\varepsilon_{r_{0}}=r_{0}/r_{m}$. Therefore, integrating the azimuthal shift function, according to Eq. (\ref{DeflectionAngle}), we can obtain an approximate expression for the light deflection angle up to second order terms proportional to $\mathcal{O}(\varepsilon_{r_{0}})$.

In order to present the light deflection angle in terms of the photon integrals of motion $L$ and $E$, we are going to calculate a power series expanssion in terms of the light ray impact parameter $b$. Since in the weak field limit the light deflection angle is small, the photons distance of the closest approach $r_{m}$ is of the same order as the impact parameter $b$. Therefore, under assumption that the wormhole throat $r_{0}\gg r_{m}$, we introduce the expansion parameter
\begin{equation}\label{epspm}
    \epsilon_{r_{0}}=\frac{r_{0}}{b}.
\end{equation}
In the frame of the given approximation, one can easily solve the integral \eqref{DeflectionAngle} in the second order terms to find the following result
\begin{equation}\label{WSDA1}
\begin{split}
	\hat{\alpha}(b) & \simeq \frac{2(1-a)r_{0}}{b} +\frac{\pi(11 - 22 a + 15 a^2)r_{0}^2}{16 b^2} \\ & +\mathcal{O}(\epsilon_{r_{0}}^3).
\end{split}
\end{equation}

Up to the first formal order in the expansion parameter the approximate expression for the light deflection angle (\ref{WSDA1}) coincides with the result \eqref{WSDAGBT} found by GBT.

In the case of the wormhole with unbounded mass function \eqref{WSUMF} one can show that the impact parameter of the light ray is given by
\begin{equation}
	b(r_{_{m}})=\frac{r_{m}}{1+(1-a)\sqrt{\frac{r_{0}}{r_{m}}}}
\end{equation}

Considering a light rays with distance of closest approach $r_{m}\gg r_{0}$, where $r_{0}$ is the wormhole throat, one can show that the light deflection angle in the weak deflection limit is given by
\begin{align}\label{WDAPWH2}
	\nonumber \hat{\alpha}(b) & \simeq - \frac{\sqrt{\pi}}{2}\frac{\Gamma\left({\frac{3}{4}}\right)}{\Gamma\left(\frac{5}{4}\right)}\sqrt{\frac{r_{0}}{b}}(1-2a) + \frac{(3-6a)r_{0}}{b} \\ & + \mathcal{O}(\epsilon_{r}^3,a^2).
\end{align}
Here, we have choosen $\epsilon_{r}=\sqrt{r_{0}/b}$ for a power series expansion parameter and we have also assumed $a\ll 1$.

Comparison of the result with that obtained by GBT \eqref{WSDA2} shows that the both expresions for the deflection angle match up to the leading term proportional to $\mathcal{O}(\epsilon_{r})$. Moreover, one can notice that the light deflection angle by wormhole with an unbounded mass function is negative, in the limit $r_{0}/b \rightarrow 0$, for small enough $a$.

\begin{figure}[t!]
            \centering
            \includegraphics[width=0.483\textwidth]{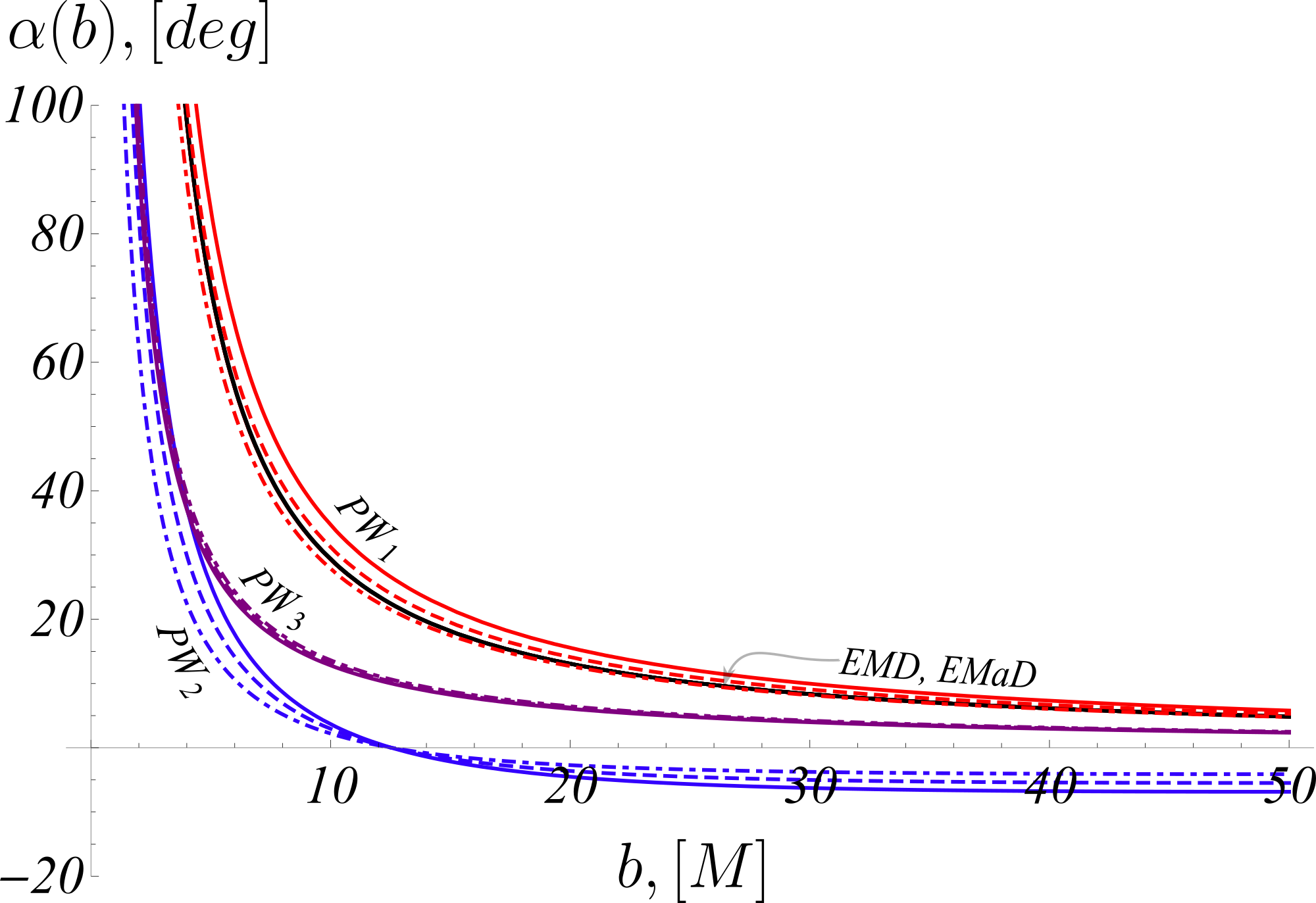} \\[1mm]	
            \caption{\label{fig:1}\small \textit{The deflection angles by EMD and EMaD black holes (black), phantom wormhole with a bounded mass function (red), phantom wormhole with an unbounded mass function (blue) and phantom wormhole with vanishing redshift function (purple) as a function of the light ray impact parameter $b$. In the cases of the EMD and EMaD black holes the values of $Q/M$ and the metric parameter $\gamma$ are the same as on Fig. \ref{fig:1}. In the cases of the phantom wormholes three different values of the metric parameters are involved as follows: wormhole with a bounded mass function: $a=-0.2$ (thick), $a=-0.1$ (dash) and $a=0$ (dash--dot); wormhole with an unbounded mass function: $a=0$ (thick), $a=0.1$ (dash) and $a=0.2$ (dash--dot); wormhole with vanishing redshift function: $\alpha=0$ (thick), $\alpha=0.1$ (dash) and $\alpha=0.2$ (dash--dot).}}\label{fig:2}
\end{figure}

In the last case concerning a wormhole with vanishing redshift function \eqref{WSVRF} one can show that the impact parameter $b$ and the closest approach distance of the light ray $r_{m}$ coincide
\begin{equation}\label{IPWH3}
	b(r_{m})=r_{m}.
\end{equation}
Following the same procedure applied in the previous cases one can show that the approximate light deflection angle is given by
\begin{equation}\label{WS3DA}
\begin{split}
	\hat{\alpha}(b) & \simeq \frac{(1+\alpha[1-\ln{2}])r_{0}}{b}+\frac{3\pi r_{0}^2}{16b^2}(1-\alpha[1-\ln 4]) \\ & + \mathcal{O}(\epsilon_{r_{0}}^2, \alpha^2),
\end{split}
\end{equation}
where first we have adopt $\epsilon_{r_{0}}=r_{0}/b$ for the series expansion parameter, as we assume that the wormhole throat $r_{0}$ and the impact parameter $b$ of the photon obey the relation $r_{0}\ll{b}$. Afterwards, we perform a power series expansion of the obtained result, assuming that the metric parameter $\alpha$ is small quantity.

Up to the first formal order in the expansion parameter the approximate expression for the light deflection angle (\ref{WS3DA}) coincides with the result \eqref{WSVRFDA} found by GBT.

The perturbation series for the deflection angles by phantom wormholes (\ref{WSDA1}), (\ref{WDAPWH2}) and (\ref{WS3DA}) are plotted on Fig. \ref{fig:2} together with the deflection angles by EMD and EMaD black holes (\ref{DAGHS}) and (\ref{DAEMaD}) for different values of the metrics parameters. In general, all the deflection angles are monotonically decreasing functions of the impact parameter $b$, with an exception of the deflection angle by the phantom wormhole with an unbounded mass function. The deflection angle of that wormhole has a negative local minimum for some $b$, and becomes a positive function for relatively small impact parameters when the weak deflection limit approximation breaks down. The deflection angle by phantom wormhole with a bounded mass function is greater than the deflection angle by the phantom black holes, while the bending angle by phantom wormhole with vanishing redshift function remains smaller. This behaviour of the deflection angles is critical to creating images by the examined candidates for gravitational lenses, as one can see from the discussion in the Sec. \ref{sec:VI}.

\section{Observables}\label{sec:VI}

In this section we pay attention on the observational relevance of the results obtained in the previous sections knowing the light deflection angles by the black holes and wormholes under consideration. To study the gravitational lensing effects in the weak deflection limit we adopt the configuration where the massive compact object (the lens $L$) is situated between a point source of light ($S$) and an observer ($O$), and the both are located in the asymptotically flat region of the spacetime, at distances much larger than the black hole horizons $r_{+}$ and $r_{-}$, as well as the wormhole throats $r_{0}$. In order to calculate the positions of the images of greatest importance to observations, we examine especially the diameter angles of the Einstein rings, which for a point source have infinite brightness \cite{blackholes-wormholes1}. To do that, we choose the Ohanian lens equation \cite{Ohanian}, which is the closest relative to the exact lens equation, because it contains only the asymptotic approximation, concerning the observer and source positions with respect to the compact object, as it is not involved in any additional assumption \cite{Bozza}. It can be written in terms of the observational angular coordinates, namely the image position $\theta$, the source position $\beta$ and the light deflection angle $\alpha$ as follows
\begin{equation}\label{LensEq}
    \arcsin\left(\frac{D_{OL}}{D_{LS}}\sin{\theta}\right)-\arcsin{\left(\frac{D_{OS}}{D_{LS}}\sin{\beta}\right)}=\hat{\alpha}(\theta)-\theta,
\end{equation}
where the $D_{OL}$ is the observer--lens distance, $D_{OS}$ is the observer--source distance and $D_{LS}$ is the lens--source distance. Here the deflection angle $\hat{\alpha}$ is a function of $\theta$ through the light ray impact parameter $b=D_{OL}\sin{\theta}$. The general solution of the lens equation (\ref{LensEq}) is \cite{Bozza}
\begin{widetext}
\begin{equation}\label{ImagePositions}
    \theta_{\pm}(\hat{\alpha},\beta)=\arctan\left(\frac{D_{OS}\cos{\hat{\alpha}}\sin{\beta}\pm\sin{\hat{\alpha}}\sqrt{D_{LS}^2-D_{OS}^2\sin^{2}{\beta}}}{D_{OL}-D_{OS}\sin{\hat{\alpha}}\sin{\beta}+\cos{\hat{\alpha}}\sqrt{D_{LS}^2-D_{OS}^2\sin^{2}{\beta}}}\right),
\end{equation}
\end{widetext}
where solutions $\theta_{\pm}$ represent the positions of the two weak field images situated in diametrically opposed directions with respect to the center of the lens. An observer, located on the optical axis passing through the lens typically sees the images located on the left $\theta_{-}(\beta<0)$ and on the right $\theta_{+}(\beta>0)$ side of the lens, according to the position of the light source $\beta$ with respect to the optical axis.

In the special situation $\beta=0$, when the source lies on (or passes through) the optical axis an Einstein ring is formed. In this particular case the lens equations is reduced to a simpler equation with solutions
\begin{eqnarray}\label{EinsteinRing}
    \theta_{+}=-\theta_{-}=\arctan\left(\frac{D_{LS}\sin\hat{\alpha}}{D_{OL}+D_{LS}\cos{\hat{\alpha}}}\right),
\end{eqnarray}
which in the weak deflection approximation, $\hat{\alpha} \ll 1$ represents the angular radius of the Einstein ring given by
\begin{eqnarray}\label{EinsteinRing1}
    \theta_{E}\simeq\frac{D_{LS}}{D_{OS}}\hat{\alpha}(b).
\end{eqnarray}
Here is taken into account that $D_{OS}=D_{OL}+D_{LS}$, when the angular source position is $\beta=0$.

In order to make comparison between the observational radii of the Einstein ring created in the black hole and wormhole scenario we are plotted in the paper for different values of the charge $Q/M$ and the metric parameter $\gamma$ and under the following assumptions. We consider the massive dark object Sgr A$^{*}$ in the center of our Galaxy as a lens. We are taking into account that the observer is positioned at distance $D_{OL}=8.33$ kpc from the lens, while for the lens--source distance, following \cite{VirbhadraD}, we have taken $D_{LS}=0.5 D_{OL}$. According to \cite{Gillessen} the lens mass is $M=4.31\times 10^{6}M_{\odot}$, so $M/D_{OL}\approx 2.49\times 10^{-11}$.

\begin{figure}[t!]
            \centering
            \includegraphics[width=0.42\textwidth]{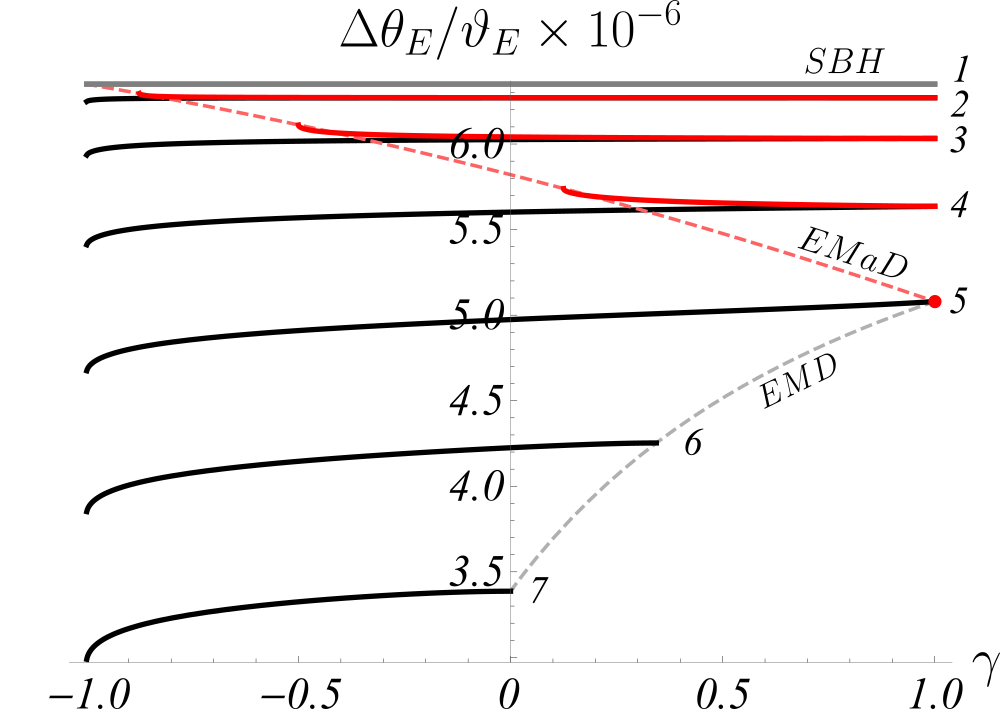} \\[1mm]	
            \caption{\label{fig:1}\small \textit{The relative angular position of the weak field Einstein ring $\theta_{E}$ for EMD (black) and EMaD (red) black holes for the following values of $Q/M$: $Q/M=0$ corresponding to the Schwarzschild black hole (curves 1), $Q/M=0.25$ (curves 2), $Q/M=0.5$ (curves 3), $Q/M=0.75$ (curves 4), $Q/M=1$ (curves 5), $Q/M=1.22$ (curves 6) and $Q/M=\sqrt{2}$ (curves 7).}}\label{fig:3}
\end{figure}

\begin{figure*}[t!]
            \centering
            \includegraphics[width=0.99\textwidth]{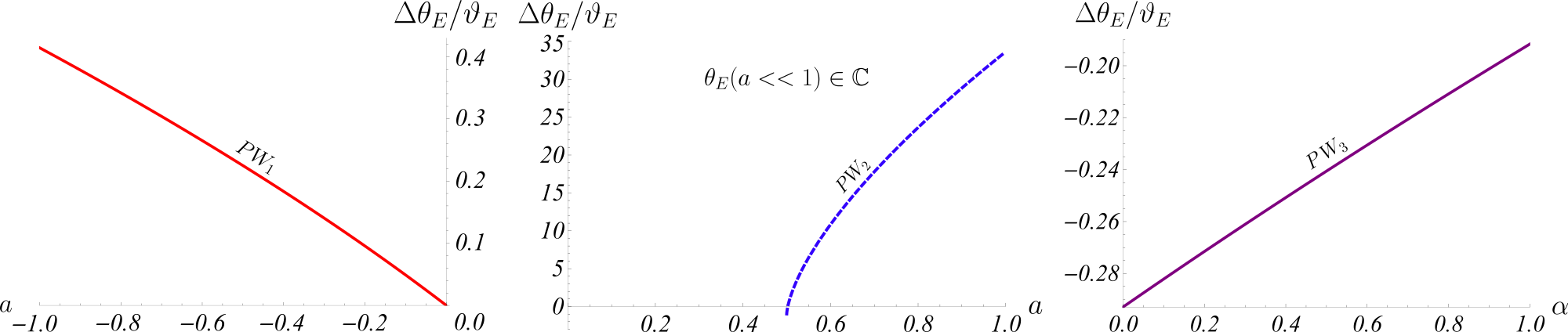} \\[1mm]	
            \caption{\label{fig:1}\small \textit{The relative angular positions of the weak field Einstein ring $\theta_{E}$ for the phantom wormhole with bounded mass function, phantom wormhole with an unbounded mass function and phantom with a vanishing redshift function versus the specific metric parameters. The shown real solutions of the lens equations are relevant only for small values of the metric parameters.}}\label{fig:4}
\end{figure*}

As a starting point we calculate the classical radius of the Einstein ring for a point-like gravitational lens based on the Schwarzschild geometry, in order to define a characteristic angular scale in the celestial sky. Keeping only the first order of the Schwarzschild black hole deflection angle $\hat{\alpha}(b)=4M/b$, and using the relation $b=D_{OL}\sin{\theta}\simeq D_{OL}\theta$ the bending angle in the hypothesis of small angles is
\begin{eqnarray}
    \hat{\alpha}_{\rm Sch}(\theta)\simeq\frac{4M}{D_{OL}}\frac{1}{\theta}.
\end{eqnarray}

Therefore, the typical Einstein ring radius of the lens system in the weak deflection limit, according to the Eq. (\ref{EinsteinRing1}), is given by
\begin{eqnarray}
    \vartheta_{E}=\sqrt{\frac{4M}{D_{OL}}\frac{D_{LS}}{D_{OS}}}\simeq1.186 \, arcsec.
\end{eqnarray}

In this situation the corresponding light deflection angle is $\hat{\alpha}_{\rm Sch}(\vartheta_{E})\simeq 988.2 \, \mu arcsec$.

Calculating the typical Einstein ring for the Schwarzschild point--like lens one can study the specific Einstein rings for the compact objects under consideration and present their sizes in units of it. Let us first consider a couple of black hole solutions with canonical electromagnetic field EMD and EMaD. Taking into account the approximate expressions for the deflection angles (\ref{DAGHS}) and (\ref{DAEMaD}) and Eq. (\ref{EinsteinRing1}) as well, we obtain the lens equation whose maximal real solution represents the weak field Einstein ring. In explicit form, the lens equation is reducing to third order algebraic equation in terms of $\theta$. The obtained positions of the Einstein ring for the different values of the electrical charge $Q/M$ and the metric parameter $\gamma$ are plotted in Fig. \ref{fig:3}. As we can see the role of the parameter $\gamma$ responsible for the coupling between the dilaton and the Maxwell field is to increase the deflection angle in a regime of weaker coupling when $\gamma$ is bigger. Besides the weaker coupling between the phantom scalar field and canonical Maxwell field leads to decreasing of the light deflection angle. In all of those cases, the increasing of the electrical charge leads to decreasing of the deflection angle and to creating of smaller in size Einstein ring. As in those situations, the deflection angle by EMD black hole is smaller than the bending angle by EMaD black hole, the created Einstein ring in the dilaton case, can't exceed the size of Einstein rings around the black hole with phantom dilaton field.

Since the differences of the created weak field Einstein rings by the dilaton and phantom dilaton black holes are negligible the probability for discriminating between these types of compact objects by observation of the rings remains very small.

In comparison with the strong deflection limit the created relativistic Einstein rings have similar behaviour with the increase of the electrical charge $Q/M$ as opposed to the impact of the coupling parameter $\gamma$ to the observables. In that situation, the weaker coupling between the canonical Maxwell and scalar fields leads to the formation of Einstein ring bigger in size than the size of the relativistic ring inherent for EMaD black hole with the same strength of the coupling between the canonical Maxwell and phantom scalar field. In both  black hole cases the increasing of the electrical charge $Q/M$ leads to negligible decreasing of the size of the relativistic rings, as they remain smaller than the Schwarzschild black hole ring \cite{Gyulchev:2012ty}.

Let us consider the phantom wormhole cases for comparison. Unlike the Maxwell-dilaton and Maxwell-phantom dilaton black hole cases, the phantom wormholes give an opportunity for more clear distinguishing of their weak field Einstein rings. Taking into account the approximate expressions for the deflection angles (\ref{WSDA1}) and (\ref{WDAPWH2}) and Eq. (\ref{WS3DA}) as well, we obtain  the lens equation whose maximal real solution represents the weak field Einstein ring. In the phantom wormhole case with a bounded mass function, the solution is greater than the solution for the Schwarzschild black hole case for non zero values of the metric parameter $a$, as it is shown in Fig. \ref{fig:4}. This indicates that there is the possibility of forming a larger-sized weak field Einstein ring than the Schwarzschild one.

In the case of the phantom wormhole with an unbounded mass function, the lens equation does not possess a real solution for small enough metric parameter $a$ which indicates the inability for the creation of weak field Einstein ring. The only possible real root of the lens equation in the given approximation exist for relatively smaller light impact parameters, for which the weak deflection limit methodology gives a huge error.

In the last case concerning the phantom wormhole with a vanishing redshift function, the lens equation owns one single positive root, which is an indication that the wormhole under consideration is capable to create a weak field Einstein ring. In this case, the ring is smaller in size than the size of the Einstein ring inherent for the Schwarzschild black hole as with the increase of the metric parameter $\alpha$ the ring that has been created slightly increase in size.

\begin{figure}[t!]
            \centering
            \includegraphics[width=0.483\textwidth]{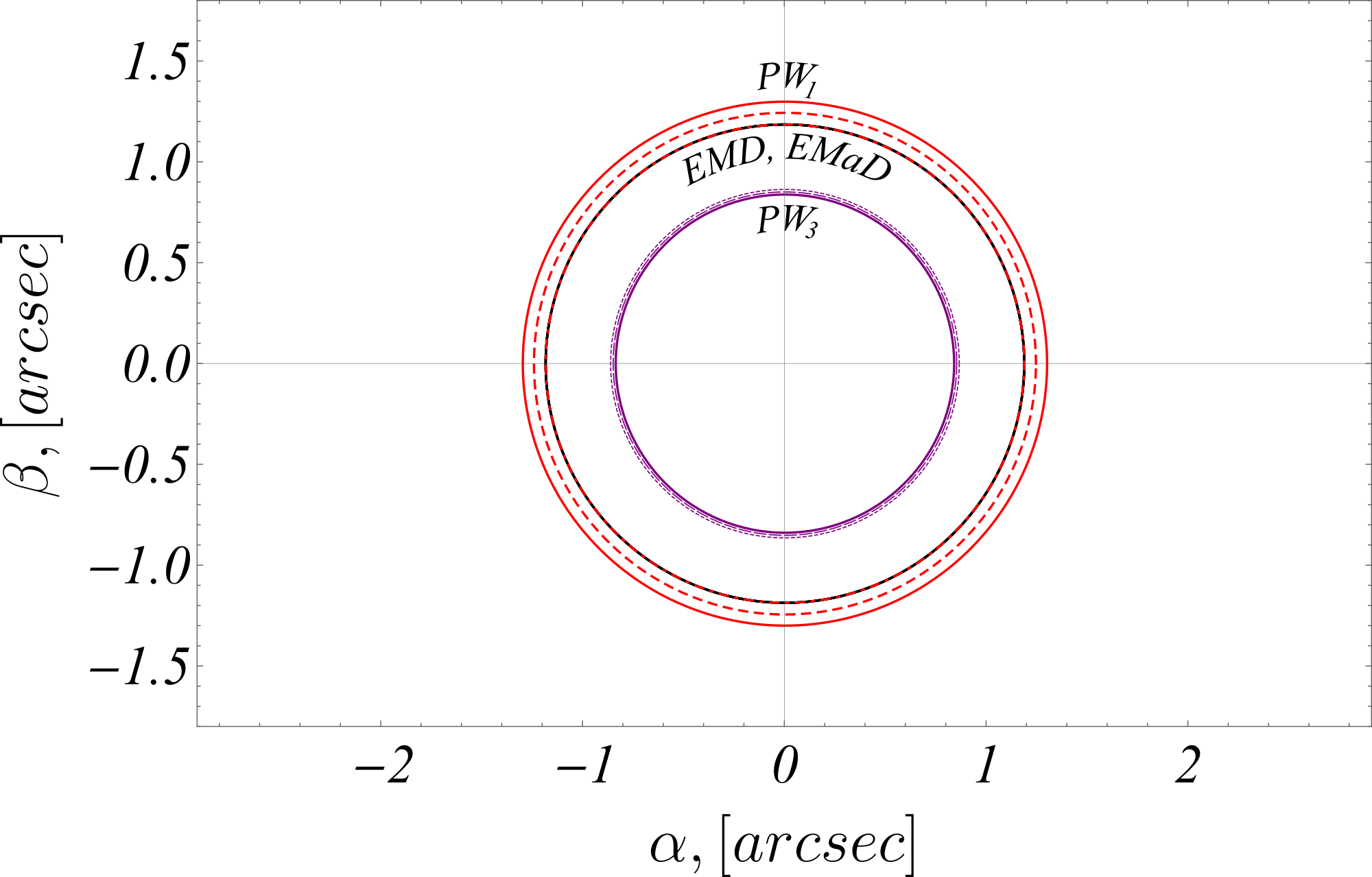} \\[1mm]	
            \caption{\label{fig:1}\small \textit{The positions of the weak field Einstein rings by EMD and EMaD black hole lenses (black), as well as the phantom wormhole with bounded mass function (red) and phantom with a vanishing redshift function lenses (purple). The phantom wormhole with an unbounded mass function does not create a weak field Einstein ring, because of the negative deflection angle at the asymptotic infinity. Here $\alpha$ and $\beta$ are the angular celestial coordinates in the observer's sky. }}\label{fig:5}
\end{figure}

The positions of the created weak field Einstein rings by the considered compact objects with exotic geometries are shown in Fig. \ref{fig:5} together with the weak field Einstein ring specific for the Schwarzschild black hole for comparison.

The results obtained in the section could be used in observational contexts to test the theories of gravity on which the
considered black hole and wormhole solutions are based. If the gravitational lens is identified with a black hole, we can estimate the presence of an electrical charge $Q/M$ or in the case of a wormhole to calculate the radius of the wormhole throat $r_{0}$ in terms of the specific for the lens system distances $D_{OS}$, $D_{OL}$ and $D_{LS}$.

\section{Conclusions}

In this paper, first we have calculated the deflection angle by GHS BH and  EMaD BH in weak field approximation. We applied the GBT to the corresponding optical geometries that provide us to calculate the deflection angle by integrating over a domain outside the impact parameter. Notice an interesting fact: the effect of the bending of light ray is a global effect.
The deflection angle by GHS BH and then by EMaD BH are calculated as follows:
\begin{equation}
\hat{\alpha}_{GHS}\simeq 2\,{\frac {{\it r_+}}{b}}+ 2\,{\frac {\gamma\,{\it r_-}}{b}}+\mathcal{O}(r_{+}^{2},r_{-}^{2}),
\end{equation}
and
\begin{equation}
\hat{\alpha}_{EMaD}\simeq 2\,{\frac {{\it r_+}}{b}}+ 2\,{\frac {{\it r_-}}{b\gamma}}+\mathcal{O}(r_{+}^{2},r_{-}^{2}).
\end{equation}
The difference between the deflection angles are plotted
in Fig. 1. As the distance of the closest approach $r_{0}$ decreases the impact parameter $b$ decreases too and the deflection of the light trajectory grows more and more.

In the case of GHS BH and EMaD BH, it is interesting to point out that, up to the first order terms, the deflection angles are reduced to the Einstein deflection angle $\hat{\alpha}\simeq 4M/b$ and are not affected at all by the electric charge $Q$ which is encoded in $r_{\pm}$. In other words, the effect of electric charge becomes significant if we take into account higher order terms.

Next, to elucidate observational differences between the deflection angle by the black hole and wormhole phenomena, the deflection angle by phantom wormholes such as wormholes with a bounded mass function, wormholes with an unbounded mass function and wormholes with vanishing redshift function are calculated by using the GBT as follows:

\begin{equation}
\hat{\alpha}_{pw_{1}}\simeq \frac{2\,r_0}{b}\left(1-a   \right)+\mathcal{O}(r_0^2,a^2),
\end{equation}

\begin{eqnarray}
\hat{\alpha}_{pw_{2}} &\simeq & - \frac{\sqrt{\pi}}{2}\frac{\Gamma\left({\frac{3}{4}}\right)}{\Gamma\left(\frac{5}{4}\right)}\sqrt{\frac{r_{0}}{b}}(1-2a)+\mathcal{O}(r_0,a^2)
\end{eqnarray}

and

\begin{equation}
\hat{\alpha}_{pw_{3}}\simeq \frac{(1+\alpha[1-\ln{2}])r_{0}}{b}+\mathcal{O}(r_0,\alpha^2).
\end{equation}

We see that besides the effect of the wormhole geometry on the bending of light encoded in $r_0$, the parameters $\alpha$ and $a$ also affects the value of the deflection angle in an interesting way. Neglecting the effect of $\alpha $ and $a$, we see that the geometric contribution depends on the particular choose of the mass function. For instance, in the case of wormhole with a bounded mass function we find a magnitude of the banding angle twice as $r_0/b$. Moreover, in the case of wormhole with an unbounded mass function the magnitude of the deflection angle can be negative at a distance far enough from the wormhole throat, where the weak field approximation is valid. Finally we should point out that, according to the perturbative analyses performed in the present paper the agreement between the deflection angles obtained via the GBT and geodesic approach is exact up to the first order terms, thus going to higher order terms the agreement breaks down. The main reason for this is the straight line approximation involved in the integration domain. In principle, this inconsistency can be resolved by choosing a more precise relation for the light ray trajectory in the integration domain. Nevertheless these phantom wormholes are interesting from a theoretical point of view because of arising in general relativity.

Afterwards, we checked these results using the geodesics method and confirmed them. In Fig. 1 we have seen that, the approximated deflection angles by GHS and EMaD black holes \eqref{DAGHSro} and \eqref{DAEMaDro} calculated by the geodesics method have a perfect coincidence with the numerically calculated deflection angles in the weak field limit $r_{\pm}/b\rightarrow 0$. Notice, that the approximation breaks down when the minimal distance of the closest approach $r_{m}$ converge to the photon sphere radius of the black holes. Moreover, the calculated approximate expressions for the deflection angles describe correctly the slightly repulsive effect on the light rays deviation by the EMaD black hole versus GHS black hole, as the numerical calculations shown.

Finally, we have study the observational relevance of the results obtained in the paper knowing the light deflection angles of the considered phantom black holes and wormholes. Assuming, the compact dark massive object at the centre of the Galaxy is a lens we have found that the wormhole with a bounded mass function can form weak field Einstein ring with larger angular radius with respect to the radius of the Einstein ring by Garfinkle-Horowitz-Str\"{o}minger dilation and Einstein-Maxwell anti-dilaton black holes, as well the Einstein ring by phantom wormhole with a vanishing redshift function.

\acknowledgments
This work was supported by the Chilean FONDECYT Grant No. 3170035 (A\"{O}).

\end{document}